\documentclass[10pt,conference]{IEEEtran} 
\IEEEoverridecommandlockouts
\usepackage{cite}
\usepackage{amsmath,amssymb,amsfonts}
\usepackage{algorithmic}
\usepackage{graphicx}
\usepackage{textcomp}
\usepackage{xcolor}

\usepackage{textcomp}
\usepackage{xcolor}
\usepackage{hyperref}
\usepackage{cleveref}
\usepackage{xspace}
\usepackage{paralist}
\usepackage{multirow}
\usepackage{subcaption}
\usepackage{tikz}
\usepackage{courier}
\usepackage{paralist}

\usepackage{arydshln}

\usepackage{comment}
\def\BibTeX{{\rm B\kern-.05em{\sc i\kern-.025em b}\kern-.08em
    T\kern-.1667em\lower.7ex\hbox{E}\kern-.125emX}}
\begin{document}

\newcommand{\company}{Microsoft\xspace}
\newcommand{\etal}{\emph{et al.}\xspace}
\newcommand{\ie}{\emph{i.e.,}\xspace}
\newcommand{\eg}{\emph{e.g.,}\xspace}

\newcommand{\bc}[1]{\textcolor{black}{#1}}

\title{Recommending Root-Cause and Mitigation Steps for Cloud Incidents using Large Language Models}

\author{
\IEEEauthorblockN{Toufique Ahmed\IEEEauthorrefmark{1}\textsuperscript{\textsection}, Supriyo Ghosh\IEEEauthorrefmark{2}, Chetan Bansal\IEEEauthorrefmark{2}
\\ Thomas Zimmermann\IEEEauthorrefmark{3}, Xuchao Zhang\IEEEauthorrefmark{2}, Saravan Rajmohan\IEEEauthorrefmark{2}}
\IEEEauthorblockA{\IEEEauthorrefmark{1}\textit{UC Davis}}
\IEEEauthorblockA{\IEEEauthorrefmark{2}\textit{Microsoft}}
\IEEEauthorblockA{\IEEEauthorrefmark{3}\textit{Microsoft Research}}
}

\maketitle

\begingroup\renewcommand\thefootnote{\textsection}
\footnotetext{This work is done during the author’s internship at Microsoft Research.}
\endgroup

\thispagestyle{plain}
\pagestyle{plain}

\begin{abstract}
Incident management for cloud services is a complex process involving several steps and has a huge impact on both service health and developer productivity. On-call engineers require significant amount of domain knowledge and manual effort for root causing and mitigation of production incidents. Recent advances in artificial intelligence has resulted in state-of-the-art large language models like GPT-3.x (both GPT-3.0 and GPT-3.5), 
which have been used to solve a variety of problems ranging from question answering to text summarization. In this work, we do the first large-scale study to evaluate the effectiveness of these models for helping engineers root cause and mitigate production incidents. We do a rigorous study at \company{}, on more than 40,000 incidents and compare several large language models in zero-shot, fine-tuned and multi-task setting using semantic and lexical metrics. Lastly, our human evaluation with actual incident owners show the efficacy and future potential of using artificial intelligence for resolving cloud incidents.
\end{abstract}


\section{Introduction}
\label{intro}


Large IT enterprises such as Amazon, Google, Microsoft, and Salesforce have replaced the traditional shrink-wrapped software and moved towards deploying applications and services on cloud platforms. In today's cloud systems, production incidents (e.g., outage or performance degradation, unplanned interruptions) adversely impact the customers and can be expensive in terms of penalty associated with service level agreement violations and engineering efforts required to mitigate the incidents. 
For example, one hour of downtime is estimated to cost Amazon US\$100 million on major shopping days~\cite{amazon-100-million}.
Despite continuous reliability efforts over the years, cloud services still experience inevitable severe incidents.

Artificial Intelligence (AI) for IT Operations, also known as AIOps, has increased in popularity. Data-driven and AI techniques have been leveraged for automating parts of the incident life-cycle, for example, incident prioritization~\cite{chen2020incidental}, retrieval of incidents with similar symptoms~\cite{saha2022mining}, and reducing the time to mitigate incidents~\cite{chen2019continuous,jiang2020mitigate}. 
However, on-call engineers (OCEs) still spend a significant amount of manual toil through multiple rounds of back and forth communication for identifying \emph{root causes} and \emph{mitigation steps}. Motivated by the recent successes of leveraging GPT-3 models for non-trivial tasks~\cite{wei2022chain,brown2020language} and code generation~\cite{chen2021evaluating}, we  apply such models to incident management. We identified the following two scenarios:
\begin{enumerate}
    \item \textbf{Find the incident's root cause.} Diagnosing incidents typically requires significant time and communication before engineers can identify the root cause of the incident. We investigate how effective large language models are at suggesting root causes for incidents (RQ1).
    \item \textbf{Suggest the mitigation steps for the incident.} After a root cause has been located, engineers take actions to mitigate the problem. We investigate how effective large language models are at recommending the mitigation steps for incidents (RQ2).
\end{enumerate}

When applying large language models several considerations and decisions need to be taken. Since the models were not trained with incident management data, is \emph{fine-tuning} of the models necessary (RQ3)? Is it more effective to build one model for each task (\emph{single-task}) or one combined model that supports both root causes and incidents (\emph{multiple task}) (RQ4)? Does the root cause help language models to find better mitigation steps (RQ5)? Do the models perform better for certain types of incidents (RQ6)? We address these questions with a rigorous large-scale evaluation of 44,340 incidents from 1,759 services of \company{}. In addition to lexical and semantic evaluation metrics that are typically reported for such experiments, we present the results from a human validation, where we asked incident owners to assess the correctness and readability of suggested root causes and mitigation steps. The original incident owners are the most qualified to assess the performance of the models on incidents. In this paper, we make the following contributions:
\begin{enumerate}
    \item This is the first work to demonstrate the usefulness of state-of-the-art large language models (LLMs) such as GPT-3.x (both GPT-3.0 and GPT-3.5)
    for resolving production incidents in a real world setting. (Section~\ref{method}) 
    \item We present a rigorous and large-scale study in \company{} on over 40,000 incidents from 1000+ cloud services with six semantic and lexical metrics. (Section~\ref{result})
\begin{itemize}
\item Fine-tuning significantly improves the effectiveness of LLMs for incident data.
\item GPT-3 and GPT-3.5 models significantly outperform encoder-decoder models in our experiments.
\item Metrics such as BLEU-4 are useful to measure relative performance of models in different settings. However, manual inspection and validation with experts is needed to assess the actual performance.
\end{itemize}
    
	\item Our human study with the actual incident owners of production incidents helps prove the efficacy of the proposed approach. (Section~\ref{hstudy}) 
\end{enumerate}


\section{Overview}
\label{overview}

\subsection{Incident management}
Production incidents are inevitable in large-scale cloud services and often severely affect the customer experience. Also, they can be extremely expensive in terms of engineering resources required to root cause and mitigate them. An incident life-cycle typically has the following four stages: 
(1) \textbf{Detection:} The first step in the incident life-cycle is detection where the incidents are either reported by internal or external customers of a given service after they notice anomalous behavior. Also, incidents can also be reported via automated monitors which are configured by the service owners. 
(2) \textbf{Triaging:} Once an incident is reported, a team of OCEs analyze the problem and route the incident ticket to appropriate engineering team. This process is often referred as incident triaging. 
(3) \textbf{Diagnosis:} The incident diagnosis and root cause identification process requires multiple iterations of back and forth communication between engineers inspecting the different aspects to understand the broad nature of the incident and identify the root cause.
(4) \textbf{Mitigation:} Based on the identified root causes, actions are taken to mitigate the problem so as to recover the service health and minimize the impact on the service users.

Lately, AIOps (AI for IT Operations) has gained popularity for automating various parts of the incident life-cycle by combining data-driven and AI techniques with data-sources like application logs, time series performance metrics and service traces~\cite{chen2019continuous,chen2020incidental,jiang2020mitigate,chen2020towards}. Albeit significant efforts, incident management in large cloud systems still requires a huge amount of engineering effort and cost. More specifically, even with plethora of historical incident data, root cause identification and mitigation remains notoriously challenging and time consuming tasks. In this work, we propose to use large language models such as GPT-3.x to automatically recommend root causes and mitigation for new incidents by leveraging historical incident data. 


\subsection{The promise of LLMs/GPT-3.x models}
Large language models (LLMs) such as GPT-3.x \cite{brown2020language} have emerged as one of the hottest trends in natural language processing over the last few years. With 175 billion parameters, the GPT-3.x language models, which held the record for being the largest neural network ever developed, is an order of magnitude larger than prior language models. Using this massive model architecture, GPT-3.x were trained using almost all accessible data from the Internet, including CommonCrawl \cite{commoncrawl}, WebText \cite{kulkarni2009collective}, Wikipedia~\cite{wikipedia}, and a corpus of books. GPT-3.x models surpass the state-of-the-art models in a variety of NLP tasks, including machine translation, question-answering, and close tasks. Furthermore, the GPT-3.x models achieved a significant milestone by showing that unsupervised language models trained with adequate data can multi-task to the same level of fine-tuned models using just a few examples of the new tasks. As a result of its powerful text generation capabilities in new tasks, GPT-3.x are used in a wide range of categories and industries, from productivity and education to creativity and gaming. For instance, GPT-3.x are used to produce creative writing, including blog posts, advertisements, and poetry, that mimics the literary style of well-known writers like Shakespeare.

\subsection{Root-causing and mitigating incidents}
Incident root-causing and mitigation is a complex process which requires significant amount of manual effort and, also, domain knowledge about the services. Incidents can be caused by various kind of issues such as code bugs, dependency failures, infrastructure issues, configuration bugs, etc. Due to the vast number of possibilities, it is non-trivial for the OCEs to root cause the incidents. Similarly, once the root cause is identified, there can be various mitigation steps which can be taken such as code rollback, hotfix, infrastructure changes, configuration update, etc. Identifying the correct mitigation step is again non-trivial and requires domain knowledge and experience. Human errors in root causing or mitigation of incidents results in not just more effort and human toil but also impact on the customers and the revenue. Fig. \ref{example} shows a real incident from a service where we can see the title and summary provided by the customer along with the actual root cause and mitigation. 

\begin{figure}[t!]

\fbox{\begin{minipage}{0.98\columnwidth}\small
\textbf{Title:} Attach vm fails with connection timeout

\vspace{2pt}
\textbf{Summary:}
The workspace is not associated with any vnet. Customer has a vm which is already running inside a vnet. They like to attach that vm into [product  omitted]. We tried the UI and CLI route, but  still fails with same connection timeout error. Error points that it resolves to some public ip [...] 

\vspace{2pt}
\textbf{Reference root cause:}
It is not supported to attach a private vm to a public workspace directly.

\vspace{2pt}
\textbf{Reference mitigation:}
Open a task to provide better official document for customer on the topic of virtual machine.
\end{minipage}}
\centering

\caption{A sample production incident.}
\label{example}
\vspace{-0.15in}
\end{figure}

In this study, we evaluate the effectiveness of large language models like GPT-3.x and Codex for root causing and mitigating production incidents. When an incident is created, the author would specify a title for the incident and describe any relevant details such as any error messages, anomalous behavior and other details which could potentially help with resolution. Once the OCE starts investigating the incident, they might get more details by communicating with the incident author or by looking at telemetry and logs. During the course of the investigation, the OCE might often updates the incident details. For our evaluation, we use the title and the summary of a given incident at the time of incident creation as input and generate the root cause and mitigation steps. This is to ensure that we only use the information which was available to the OCE when they started investigating the incident.

\subsection{Research questions}
\label{rq}
We investigated several OpenAI GPT-3.x models (\ie Curie, Codex-cushman, Davinci, Code-davinci-002) to generate root causes and mitigation plans for the incident. This leads to several RQs. 

\smallskip
\noindent{\underline{\em RQ1} }\emph{Are fine-tuned GPT-3.x models effective at finding the incident's root cause?}

\noindent The OpenAI models are not trained with the incident management data since the data contain sensitive privacy information, and \company follows standard protocols to ensure the security of the data. Therefore, the GPT-3.x models are not expected to perform well in zero-shot/few-shot settings. In this paper, we fine-tuned four different GPT-3.x models with different capacities and observed how the models performed at proposing the root causes of the incident.   

\smallskip
\noindent{\underline{\em RQ2} }\emph{Are fine-tuned GPT-3.x models capable of suggesting the mitigation plan for the incident?}

\noindent We are also interested in generating mitigation plans for the incident using GPT-3.x models. Like root cause generation, we fine-tune and evaluate the model using the input and criteria we use for RQ1. 

\smallskip
\noindent{\underline{\em RQ3} }\emph{How much fine-tuning improves over zero-shot learning performance of GPT-3.x models?}

\noindent Though we primarily focus on fine-tuning, GPT-3.x models are reported to be effective at various downstream tasks with zero-shot and few-shot training~\cite{brown2020language,chen2021evaluating}. In few-shot learning, we use a few examples in the prompt as input to the model, and the model generates the expected output. Zero-shot is similar to few-shot training, but none of the examples are given. These two settings are economically and environmentally beneficial (reduced carbon footprint) because we are not updating any parameters of the models. This paper will investigate how the models perform at zero-shot settings. Note that few-shot learning is unsuitable for our project because we have long sequences in our dataset, and we observe the truncation of the sequences when we infer only one sequence after fine-tuning.  

\smallskip
\noindent{\underline{\em RQ4} }\emph{Does multi-task learning improve the performance of GPT-3.x models at finding root causes and mitigation plans?}

\noindent Multi-task learning is effective for some pre-trained models~\cite{wang2021codet5}. So far, we have discussed separate training models and using the input independently to generate the incident's root cause and mitigation plans. We are interested in how GPT-3.x models react to multi-task learning in our specific setting. We combine all the training data for this experiment for both tasks. However, during evaluation, we used the same test sets used in RQ1 and RQ2.  

\smallskip
\noindent{\underline{\em RQ5} }\emph{Do GPT-3.x models get better at proposing mitigation plans if the root cause is given?}

\noindent Mitigation plans for an incident depend on the specific root cause. Different root causes may lead to different mitigation plans. Moreover, the GPT-3.x models can be improved by making the input larger or more informative. We will also investigate whether providing the root cause in the input help models find the mitigation plans. 

\smallskip
\noindent{\underline{\em RQ6} }\emph{Do the models better propose mitigation plans for machine-detected incidents than human-detected ones?}

\noindent Incidents can be machine-detected (by some monitors) or human-detected. Both types of incidents have specific characteristics. Machine-detected incidents are generally triggered when the monitor observes system changes like build failures, resource availability, request counts, etc. On the contrary, human-detected incidents are unique and may apply to a specific customer (\eg webpage is not loading). In the research question, we will investigate if the model performs well for incidents belonging to a specific class.

\subsection{Human validation}
Root causes and mitigation plans can be written in different forms. Unlike natural language translation or code summarization, root causes and mitigation steps are much more open-ended. Depending on the author, the root causes and mitigation plans can vary from generic to specific. Automatic metrics may fail to reflect the overall performance of the models ideally because these metrics compare the models' suggestions with one reference, which may be completely different from the models' correct and relevant outputs. To better understand the model's performance, we went to the owner/resolver of the specific incidents and presented the solutions from our models and baselines. They assigned correctness and readability scores to the model's output. We will discuss our methodology and findings from the human validation in Section~\ref{hstudy}.            
\section{Methodology}
\label{method}

\subsection{Dataset Preparation}
Thousands of incidents with different severity are being detected (by both machines and humans) every day at \company. The on-call engineers (OCEs) are working relentlessly to provide seamless service to the customers. To manage incidents at that scale, \company has a well-designed website for reporting and managing the incident. A database also keeps track of the website's data insertion, modification, and deletion from incident reporting to mitigation. One of the inputs to the model is the summary written at the time of incident reporting or creation to prevent any data leakage from input to output. 

In most cases, the OCEs do not follow any specific format to write incident summaries, root causes, and mitigation plans. The fields, especially summaries, contain information in multiple forms, including tables, links to prior incidents, and images of individual monitor output or code snippets. This is because the incidents are very different from each other, and the utmost priority of the OCEs is to resolve the incident rather than document the symptoms. Also, some incidents are transient and auto-mitigated. No postmortem is done if the severity is low. Since GPT-3.x are text models, we discarded the tables and images from the summaries. Hence, there is a chance that we lost some critical information while discarding that information.
      
We collected data for incidents from the database that has the creation date between January 1, 2018, to July 15, 2022. Initially, we collected 123,953 instances for root causes and 23,544 mitigations from the ``Resolved'' or ``Mitigated'' incidents with severity levels 0-3 (most severe incidents belong to level 0). The samples for mitigation are low because they can be found in the postmortem of the incident, and post-mortem are not done for every incident. After collecting the data, we observe many incidents with duplicate root causes and mitigations. Some severe incidents/ denial of service trigger hundreds of incident reports for the same event, all of which have the exact root causes and mitigations. To fairly evaluate the model, we remove the exact duplicates for root causes and mitigation plans and end up with 57,520 root causes and 8,300 mitigation plans. The average root causes and mitigations lengths are 87 and 12 tokens, respectively. Some root causes are very long, and it is difficult for the models and human evaluators to generate and evaluate the models' output. We kept the root causes up to 100 tokens, allowing us to keep 73\% of the instances for root causes. We also discarded root causes and mitigation plans with less than three tokens because those are not informative.  

After deduplication and filtering, we sorted the data according to the creation date to use only historical data for training the model. We selected 35820, 3000 and 2000 root causes for training, testing and validation. We have fewer instances for mitigations. Hence, the training, test and validation sets for mitigations have 5455, 2000 and 500 data, respectively. Even after this rigorous filtering and deduplication of data, some root causes and mitigations do not carry any useful information (\eg root cause in a different link, transient, and auto-mitigated incidents). We manually went through 3000 root causes and 2000 mitigation plans from test sets and selected 2,621 root causes and 1,780 mitigation plans. \footnote{We cannot share the dataset because incident data can contain confidential and private data and sharing such data would violate the terms of service.}

\subsection{OpenAI models and baselines}
The recent advancement of the deep neural network models is greatly influenced by the introduction of Transformer models~\cite{vaswani2017attention}. Prior approaches (\ie LSTM~\cite{hochreiter1997long} and GRU~\cite{chung2014empirical}) modeled the sequential dependencies of the generated text using recurrent architecture. These recurrent models use ``Back-Propagation through Time'' (BPTT) to recursively propagate loss values over gradients within the same recurrent units prohibiting the possibility of parallel computation while capturing the long-distance dependencies of the tokens in the sequence. Bahdanau \etal introduced an attention mechanism that works on top recurrent architecture and improves the performance of recurrent neural models by providing an attention vector that indicates the relevant part of the input to the target output~\cite{bahdanau2014neural}. Transformer model completely removes the recurrence unit and entirely relies on the attention mechanism. It uses a multi-layer, multi-head self-attention architecture where the attention mechanism can relate different positions of a single sequence to compute a sequence representation. 

Pre-trained models are currently achieving state-of-the-art performance for various natural language and code tasks. These pre-trained models work in 2 stages (\ie pre-training and fine-tuning). In the pre-training stage, we train the model to learn statistics of language (or code) in a self-supervised fashion from large-scale corpora. After that, we use 
a smaller labeled dataset to fine-tune the model for specific tasks. It is nearly infeasible to have sufficient labeled data to train such high-capacity deep learning models. Pre-trained models enable us to train such big models with the unlabeled data in a self-supervised way in the pre-training stage. All the recent pre-trained (encoder-only and encoder-decoder) models (\eg BERT~\cite{devlin2018bert}, RoBERTA~\cite{liu2019roberta}, BART~\cite{lewis2019bart}, T5~\cite{raffel2019exploring}) and decoder-only generative models (\eg GPT~\cite{radford2018improving}, GPT-2~\cite{radford2019language}, GPT-3~\cite{brown2020language}, OPT~\cite{zhang2022opt}) are basically Transformer models of various capacity trained with different pre-training objectives. The following subsections briefly discuss the baselines and OpenAI models we used for our experiments.

\subsubsection{Baselines encoder-decoder models}
We can apply the encoder-decoder models for both root cause and mitigation. The encoder will encode the input, and the decoder will generate the root cause or mitigation using the encoded representation provided by the encoder. 

Pre-trained NLP models (\eg BERT~\cite{devlin2018bert}, RoBERTa~\cite{liu2019roberta}, BART~\cite{lewis2019bart}, T5~\cite{raffel2019exploring}) use different self-supervised pre-training objectives to learn robust language representations. NLP models have programming language counterparts (\eg CodeBERT~\cite{feng2020codebert}, GraphCodeBERT~\cite{guo2020graphcodebert}, PLBART~\cite{ahmad-etal-2021-unified}, CodeT5~\cite{wang2021codet5}, NatGen~\cite{chakraborty2022natgen}) where the models are initialized with the NLP models' weights and continued pre-training with code and associated natural language comments in most cases. Though root cause and mitigation are natural language descriptions, the vocabulary (\eg identifiers) overlaps more with the comments used in code models. Therefore we picked both NLP and code models from OpenAI and baseline criteria to see if the performance differs depending on the domain used for pre-training. For baselining, we pick RoBERTa~\cite{liu2019roberta} and CodeBERT~\cite{feng2020codebert} models because of two reasons: i) these two models are architecturally identical with 125M parameters,  ii) Both models are widely used as baselines (in fact, CodeBERT is the primary baseline model of the CodeXGLUE~\cite{DBLP:journals/corr/abs-2102-04664} dataset, which is a popular benchmark of 10 SE tasks including encoder-decoder tasks like code summarization and code translation). Note that many transformer-based encoder-decoder models can be applied to this problem. However, comparing with all the models is beyond the scope of the paper. 

\noindent{\underline{\bf\em RoBERTa:} }BERT is the first model that introduced the pre-training strategy that outperforms the traditional Transformer models. It applied two pre-training strategies: Masked Language Modeling (MLM) and NSP (Next Sentence Prediction). In MLM pre-training, we randomly mask out 15\% of the tokens and ask the model to recover those tokens, whereas, in NSP, we train the model to learn to predict the next sentence following an input sentence. Liu \etal \cite{liu2019roberta} propose RoBERTa (A Robustly Optimized BERT
Pre-training Approach), which outperforms the BERT model with a few changes, such as dynamic masking and dropping NSP, achieves better performance. We apply RoBERTa as NLP baseline model. 

\noindent{\underline{\bf\em CodeBERT:} }CodeBERT is architecturally identical to RoBERTa model that uses two pre-training objectives: MLM and Replaced Token Detection (RTD)~\cite{clark2020electra}. We can define RTD as a binary classification problem where two data generators (i.e., NL and PL) generate plausible alternatives for a set of randomly masked positions. A discriminator is trained to determine whether a word is the original one or not. CodeBERT is pre-trained on CodeSerachNet~\cite{husain2019codesearchnet} dataset.

\subsubsection{OpenAI generative models}
Radford \etal introduced general task-agnostic generative pre-training of language models (GPT) and outperformed 9 out of 12 discriminatively trained models that use architectures designed for the specific task~\cite{radford2018improving}. In generative pre-training, we autoregressively predict the probability of a token given the previous tokens moving from left to right. This left-to-right autoregressive training prevents the model from retrieving information from future tokens. All the subsequent generative models (\eg GPT-2, GPT-3) use very similar pre-training objectives but have a higher capacity than previous ones and are pre-trained on a much larger dataset. Very large language models (LLMs) like GPT-3.x have 175 billion parameters and are found to be effective with few-shot learning replacing the need for fine-tuning for a specific set of tasks. However, fine-tuning GPT-3.x models are still beneficial for some tasks~\cite{brown2020language}. This paper evaluates our approach using four OpenAI\cite{openai} GPT-3.x models: Curie, Codex, Davinci, and Code-davinci-002.     

\smallskip
\noindent{\underline{\bf\em Curie:} }Curie is the fastest GPT-3 model with 6.7B parameters. This model is trained with natural language data and performs well on language translation, complex classification, text sentiment, and summarization tasks. This is the smallest model we use for our experiments.  

\noindent{\underline{\bf\em Codex:} }The Codex models are also GPT-3 models trained for understanding and generating code. The training data contains both natural language and billions of lines of public code from GitHub. We use one model, Codex-cushman from Codex family, with 12 billion parameters. Though the models are pre-trained for code-related tasks, it somehow relevant to incident management. Root cause and mitigation contain a lot of terminology (\eg filenames, identifiers) which relate more to comments used in software development projects. 

\noindent{\underline{\bf\em Davinci:} }Davinci is the biggest GPT-3 model (175 billion parameters) we use for our experiments. It can perform tasks with fewer instructions than other GPT-3 models. Davinci usually performs better at understanding the content or creative content generation task. It is also very good at solving logic problems. However, training the 175 billion parameters model is costly and requires a much longer period (almost four times compared to Curie with the same dataset) and more resources. Davinci is not trained to understand or generate code.  

\noindent{\underline{\bf\em Code-davinci-002:}} Code-davinci-002 is the 175 billion parameters GPT-3.5 model we use for our experiments. Code-davinci-002 is an upgraded and more capable version of Codex model that was trained on a more recent dataset of text and code corpus. 

\subsection{Model configuration}
One of the limitations of pre-trained encoder-decoder models is that they can only encode 512 tokens. We observe that several samples from our test set truncated in GPT-3 model even though GPT-3 models support from 2048 tokens (\eg Curie, Codex) to 4000 tokens (\eg Code-davinci-002). Therefore, we can assume that the traditional encoder-encoder models do not have enough capacity to encode our sequences. 

Encoder-decoder models have been successful for problems like code summarization~\cite{feng2020codebert,ahmad-etal-2021-unified,wang2021codet5}, code translation~\cite{DBLP:journals/corr/abs-2102-04664}, and natural language translation~\cite{vaswani2017attention,lewis2019bart,raffel2019exploring}. We usually generate one sample using beam search for each input and compare the results with the reference. Generating one sample is sufficient for these problems because the target text is less open-ended. Besides, most of the information needed for successful generation can be found in the input for this set of problems. The models need to learn the syntactic alignment between two programming languages for code translation. Learning to transform conditional statements and loops from one programming language to another may be enough to do a successful translation, which is learnable from a few thousand samples. For natural language translation learning the mapping between the words from different natural languages is essential to generate good quality translation. Code summarization is slightly different from these two, where the input is much longer than the output. However, Ahmed and Devanbu found that all the necessary information for code summarization is extracted from the identifiers, and obfuscating the identifiers hurts the models~\cite{ahmed2022multilingual}. Generating root causes and mitigation plans is much more complex than these problems, where the input may not contain handy information. The models need to be able to generate more diverse and creative solutions to answer the question. Our problem is more aligned with code generation problems where the input does not carry most information. For these types of problems, it is found that instead of using the encoder-decoder model, decoder-only models (\eg GPT-3.x) are more successful where we only focus on the following tokens considering the prior tokens generated by the models. It is well-established that encoder-decoder models are not as successful as decoder-only models in code generation tasks. However, we still apply encoder-decoder models to our problems and discuss our findings in the following sections. For RoBERTa~\cite{liu2019roberta} and CodeBERT~\cite{feng2020codebert} we use the exact setup that is used for the code summarization task~\cite{CodeXGlue,husain2019codesearchnet}. We adjust the length to 512 tokens with a batch size of 8 to provide as much as information to the model.

Full fine-tuning that retrains all the parameters is very costly and challenging for the OpenAI models with billions of parameters. We use LoRA (Low-Rank Adaptation), a novel approach that significantly reduces the number of trainable parameters by freezing the pre-trained model weights and injecting trainable rank decomposition matrices into each layer of the Transformer architecture~\cite{hu2021lora}. Even though LoRA reduces trainable parameters, it performs on-par or better than fine-tuning in model quality on RoBERTa, DeBERTa, GPT-2, and GPT-3. We fine-tuned the OpenAI GPT-3 (\ie Curie, Codex, Davinci) and GPT-3.5 (Code-davinci-002) models for root causes and mitigation plans generation. We train both models for 2000 steps (4 epochs) which OpenAI recommends. For fine-tuning smaller models (\ie Curie and Codex), we use one NVIDIA V100 GPU, and for Davinci, we use four NVIDIA V100 GPUs. For finetuning Code-davinci-002 model, we use four NVIDIA A100 GPUs. We evaluated the models on the validation set after every 100 steps and chose the model that showed minimum training loss on the validation set.

As discussed earlier, the model needs to generate more diverse and creative recommendations to solve problems like the predictions of root causes and mitigation plans. Two critical parameters to control the quality of the generated outputs are \emph{temperature} and \emph{top\_p}, and it is recommended to update one parameter. Following prior works~\cite{chen2021evaluating,xu2022systematic}, we decided to update the value of temperature. Higher temperature encourages the model to take more risk, which is necessary for the creative application~\cite{openai}. Lower value performs argmax sampling, which is very similar to what we do in encoder-decoder model models like CodeBERT. Typically, a temperature between 0.50–0.90 is the most common for creative tasks. However, a high temperature is hurtful (makes the output too diverge)~\cite{xu2022systematic}. We perform a grid search and choose 0.7 for Curie, Codex, and Davinci models and 0.5 for Code-davinci-002 experiments to minimize the divergence issue for generating five samples.  

\subsection{Evaluation Metrics}
We briefly describe the evaluation metrics used for the two downstream tasks, root cause and mitigation generation.
\subsubsection{Lexical Metrics}
For lexical metrics, we employ the smooth sentence \textbf{BLEU-4} (Bilingual Evaluation Understudy) \cite{lin2004orange} metric to calculate n-grams overlap from 1 to 4 between the reference and generated texts. In addition, the Rouge metric (Recall Oriented Understudy for Gisting Evaluation) \cite{lin2004rouge} is used to compare a candidate document to a set of reference texts. Specifically, we choose \textbf{ROUGE-L}~\cite{lin2004rouge}, which takes into account sentence-level structural similarity and identifies longest co-occurring in sequence n-grams based on Longest Common Subsequence (LCS) \cite{hirschberg1977algorithms} statistics. \textbf{METEOR} (Metric for Evaluation of Translation with Explicit Ordering) \cite{banerjee2005meteor} is the final lexical metric we selected, which is based on the harmonic mean of unigram precision and recall as well as stemming and synonymy matching as extra features.

\subsubsection{Semantic Metrics}
Since the lexical metrics usually conduct exact word matches and disregard the meaning of words, we choose three semantic metrics to evaluate our outcomes according to their semantic meanings. We use the \textbf{BERTScore} \cite{zhang2019bertscore}, which leverages the pre-trained contextual embeddings from the BERT \cite{devlin2018bert} model and matches candidate and reference sentence words based on cosine similarity. Then, the \textbf{BLEURT} score \cite{sellam2020bleurt} is selected to demonstrate the degree to what extent the candidate is fluent and conveys the meaning of the reference. Last, we select \textbf{NUBIA} (NeUral Based Interchangeability Assessor) \cite{kane2020nubia}, a recent neural-based measure that incorporates the semantic similarity, logical inference and sentence legibility from exposing layers of pre-trained language models, including RoBERTa STS \cite{liu2019roberta}, RoBERTa MNLI and GPT-2 \cite{radford2019language}.

The semantic metric calculation takes significant time and requires expensive GPU resources (Tables~\ref{frc} and~\ref{fm} took two days on a single GPU). Therefore, we reported semantic metrics for the first two research questions, and for the remaining research questions, we restricted ourselves to lexical ones that are computationally less expensive.

\section{Result}
\label{result}

\subsection{How effective are fine-tuned GPT-3.x models in generating incidents' root cause recommendation? (RQ1)}

\begin{table*}[ht]

\centering
\caption{Effectiveness of fine-tuned GPT-3.x models at finding \textbf{root causes} of the incidents}
\vspace{-0.05in}

\resizebox{.8\textwidth}{!}{%
\renewcommand{\arraystretch}{1.2}

\begin{tabular}{lcccccccccccc}
\hline
\multirow{2}{*}{Model} & \multicolumn{2}{c}{BLEU-4} & \multicolumn{2}{c}{ROUGE-L} & \multicolumn{2}{c}{METEOR} & \multicolumn{2}{c}{BERTScore} & \multicolumn{2}{c}{BLEURT} & \multicolumn{2}{c}{NUBIA} \\
                       & Top1         & Top5        & Top1         & Top5         & Top1        & Top5         & Top1          & Top5          & Top1         & Top5        & Top1        & Top5        \\ \hline
RoBERTa                & {4.21}         & NA          & {12.83}        & NA           & {9.89}        & NA           & {85.38}         & NA            & {35.66}        & NA          & 33.94       & NA          \\
CodeBERT               & 3.38         & NA          & 10.17        & NA           & 6.58        & NA           & 84.88         & NA            & 33.19        & NA          & 39.05       & NA          \\ \hdashline
Curie                  & 3.40         & {6.29}        & 9.04         & 15.44        & 7.21        & 13.65        & 84.90         & {86.36}         & 32.62        & 40.08       & 33.52       & 49.76       \\
Codex                  & 3.44         & 6.25        & 8.98         & {15.51}        & 7.33        & 13.82        & 84.85         & 86.33         & 32.50        & {40.11}       & 33.64       & 49.77       \\
Davinci                & 3.34         & 5.94        & 8.53         & 15.10        & 6.67        & 12.95        & 83.13         & 84.41         & 31.06        & 38.61       & \textbf{35.28}       & 50.79 \\
\bc{Davinci-002}        & \textbf{4.24} & \textbf{7.15} & \textbf{11.43} & \textbf{17.2} & \textbf{10.42} & \textbf{16.8} & \textbf{85.42} & \textbf{86.78} & \textbf{36.77} & \textbf{42.87} & 32.3 & \textbf{51.34} \\
\bc{\%gain for Davinci-002}  & 23.26 & 13.67 & 26.44 & 10.90 & 42.16 & 21.56 & 0.61 & 0.49 & 12.72 & 6.88 & -8.45 & 1.08 \\ 
 \hline     
\end{tabular}
}
\label{frc}
\end{table*}

Table~\ref{frc} presents the effectiveness of our baseline encoder-decoder models and fine-tuned GPT-3.x models for root cause recommendation. We have 2621 test samples for evaluating the models. We generated ten samples for the OpenAI models for two reasons: i) using temperature, we can generate very diverse and creative samples from GPT-3.x models. ii) we found that GPT-3.x models can generate valuable suggestions even with lower ranks. We observed the average BLEU-4 of all the samples at a particular rank, and we found that all the OpenAI GPT-3.x models produce examples with higher BLEU-4 even at rank eight or lower. However, ten examples are too many for a human OCE, and we restrict ourselves to five top suggestions from the model. In Table~\ref{frc}, for each metric, we have Top 1 and Top 5. Top 1 presents the mean of the first candidates for all the test samples; while calculating Top 5, we take the maximum value from the first five candidates and then find the average for all samples. This Top 5 gives an overall view of how the models are performing. For our baseline encoder-decoder models, we have only one sample for each model. 

Surprisingly, the encoder-decoder models are doing really good compared to GPT-3 models in all six automatic metrics. In fact, all six metrics fail to distinguish significant differences between the OpenAI models. The reason behind the success of encoder-decoder models in automatic metrics is that these models are less explorative and try to maximize the success depending on argmax probabilities during decoding. Now ``There is a bug in the code'' is a very common and generic sentence that can be a part of any root causes. The models maximize the success just by copying that particular segment, and automatic metrics also fail here. We tried three semantic metrics to resolve that issue, but the encoder-decoder models still benefit from the automatic metric. Table~\ref{unique} presents the number of unique samples generated by the models. For OpenAI models we only consider the first candidate to make a fair comparison. We observe that the unique candidate count for RoBERTa and CodeBERT are 6.10\% and 16.67\% of the total count, whereas, for all the OpenAI GPT-3.x models, the percentages are above 97\%. Remember that we deduplicated the dataset, and repeatedly generating the same samples should not help here. In Section~\ref{hstudy}, we interviewed the incident owners, and the majority of them complained about the generic nature of encoder-decoder models' recommendations, and these models underperform at correctness criteria. Among OpenAI models, GPT-3.5 (i.e., Code-davinci-002) model significantly outperforms all GPT-3 models as well as other baselines in terms of all the 6 automated metrics.   

Though the automatic metrics fail to detect the weaknesses of the encoder-decoder models, these metrics are still widely used. Human evaluation is hard to perform in every scenario, and these metrics can be useful to find the models' relative performance. Therefore, even though we achieve a low score on these metrics, these are useful while trying to capture the relative performance of the model in different settings. Also, getting a lower score with lexical metrics is not surprising because lexical metrics only consider token overlaps and root cause and mitigation are open-ended, and the same root cause/mitigation can be written differently. In Section~\ref{hstudy}, from the interviews with OCEs, we found that suggestions with lower BLEU-4 or other metrics are still helpful.

\subsection{How effective are fine-tuned GPT-3.x models in recommending mitigation plans for an incident? (RQ2)}

Table~\ref{fm} shows that we achieved a slightly higher mitigation score (4.44-6.76 BLEU-4) than the root cause recommendation (3.38-4.24 BLEU-4). We observed a similar and consistent pattern (Table~\ref{unique}) of the output as observed with root causes. The encoder-decoder models generate generic comments (\eg ``the issue is self-mitigated'', ``fix deployed to all regions'') like before, and those recommendations are mostly useless for the OCEs. For both RQ1 and RQ2, the fine-tuned Davinci model (even with 175 Billion parameters) is significantly underperforming other baseline methods according to automatic metrics. However, the Davinci and Code-davinci-002 models are the best performing models according to the incident owners (see Section~\ref{hstudy}) 

\begin{table*}[ht]

\centering

\vspace{0.00in}
\caption{Effectiveness of fine-tuned GPT-3.x models at finding mitigation plans of the incidents}
\vspace{-0.05in}

\resizebox{.8\textwidth}{!}{%
\renewcommand{\arraystretch}{1.2}

\begin{tabular}{lcccccccccccc}
\hline
\multicolumn{1}{c}{\multirow{2}{*}{Model}} & \multicolumn{2}{c}{BLEU-4} & \multicolumn{2}{c}{ROUGE-L} & \multicolumn{2}{c}{METEOR} & \multicolumn{2}{c}{BERTScore} & \multicolumn{2}{c}{BLEURT} & \multicolumn{2}{c}{NUBIA} \\
\multicolumn{1}{c}{}                       & Top1        & Top5         & Top1         & Top5         & Top1        & Top5         & Top1          & Top5          & Top1         & Top5        & Top1        & Top5        \\ \hline
RoBERTa                                    & 4.44        & NA           & 7.10         & NA           & 4.52        & NA           & 86.33         & NA            & 26.80        & NA          & 14.90       & NA          \\
CodeBERT                                   & {6.02}        & NA           & 4.40         & NA           & 3.37        & NA           & {86.83}         & NA            & {28.44}        & NA          & {27.89}       & NA          \\ \hdashline
Curie                                      & 5.47        & 10.62        & 8.03         & {16.31}        & {6.22}        & 12.75        & 85.65         & 87.13         & 27.20        & 37.23       & 15.30       & 25.46       \\
Codex                                      & 5.53        & 10.62        & {8.15}         & 16.23        & 6.19        & {13.15}        & 85.68         & {87.35}         & 28.43        & {37.92}       & 15.77       & {26.33}       \\
Davinci                                    & 5.54        & {10.66}        & 8.10         & 15.96        & 6.08        & 12.49        & 85.72         & 87.19         & 27.15        & 37.00       & 15.71       & 25.61  \\ 
\bc{Davinci-002}        & \textbf{6.76} & \textbf{11.66} & \textbf{10.22} & \textbf{18.14} & \textbf{8.23} & \textbf{15.13} & \textbf{86.17} & \textbf{87.65} & \textbf{30.19} & \textbf{38.96} & \textbf{17.58} & \textbf{28.81} \\
\bc{\%gain for Davinci-002}  & 22.02 & 9.38 & 25.40 & 11.22 & 32.32 & 15.06 & 0.52 & 0.34 & 6.19 & 2.74 & 11.48 & 9.42
 \\ 
\hline     
\end{tabular}
}
\vspace{-0.1in}
\label{fm}
\end{table*}



\begin{table}[ht]

\centering
\caption{Uniqueness of the models' suggestions}
\vspace{-0.05in}
\label{unique}

\resizebox{.95\columnwidth}{!}{%
\renewcommand{\arraystretch}{1.2}

\begin{tabular}{lcccc}
\hline

\multicolumn{1}{c}{\multirow{3}{*}{Model}} & \multicolumn{2}{c}{Root cause}                                                                                                        & \multicolumn{2}{c}{Mitigation}                                                                                                        \\
\multicolumn{1}{c}{}                       & \begin{tabular}[c]{@{}c@{}}\# of unique   \\ recommendations\end{tabular} & \begin{tabular}[c]{@{}c@{}}In \% of\\  total\end{tabular} & \begin{tabular}[c]{@{}c@{}}\# of unique   \\ recommendations\end{tabular} & \begin{tabular}[c]{@{}c@{}}In \% of\\  total\end{tabular} \\ \hline
RoBERTa    & 160    & 6.10     & 4      & 0.22     \\
CodeBERT  & 437   & 16.67   & 2      & 0.1       \\ \hdashline
Curie    & 2612        & 99.65       & 1669   & 93.76   \\
Codex      & \textbf{2614}      & \textbf{99.73}     & \textbf{1743}    & \textbf{97.92}     \\
Davinci    & 2587  & 98.70   & 1731  & 97.24  \\ 
\bc{Davinci-002} & \textbf{2614} & \textbf{99.73} & 1696 & 95.28
  \\ 
\hline                                                   
\end{tabular}
}
\end{table}

\subsection{How much fine-tuning improves over zero-shot learning performance of GPT-3.x models? (RQ3)}

As discussed in Section~\ref{rq}, we will investigate the performance of OpenAI models in the zero-shot setting. Table~\ref{zsr} presents the performance of the OpenAI models for root cause and mitigation. As expected, the model did not perform well in this setting since the models were not trained on confidential data from the incident management space. The models achieve  0.80-2.18 BLEU-4 for the top candidate, which is much lower (210\%) than what we achieved with fine-tuning the models (5.47-6.76) while recommending mitigation steps. Though we achieved a higher score for mitigation than root cause during fine-tuning, in the zero-shot setting, the numbers for root cause are slightly high (1.18-2.83 for the top candidates).   The model tries to complete the sequence depending on the given input. Copying a few tokens from input may help the model because the root cause is usually longer than mitigation and tends to share more tokens with the input. Because of unigram overlaps METEOR is doing better compared to other metrics (BLEU-4 and ROUGE-L) because it looks for the unigram precision and recall, making it lenient compared to BLEU-4 and ROUGE-L. We observe another interesting phenomenon here. Though the Davinci model was underperforming in RQ1 and RQ2, it significantly outperforms the other OpenAI models at zero-shot settings for both root cause and mitigation. This is because the model has higher parameters and is trained on more data enabling it to infer better without explicit training.

\begin{table}[t]

\centering
\caption{Effectiveness of OpenAI models for recommending root causes and mitigation steps at zero-shot setting}
\vspace{-0.05in}
\label{zsr}

\resizebox{.95\columnwidth}{!}{%
\renewcommand{\arraystretch}{1.2}

\begin{tabular}{p{1.1cm}p{1.3cm}cccccc}
\hline
\multicolumn{1}{c}{\multirow{2}{*}{Objective}} & \multicolumn{1}{c}{\multirow{2}{*}{Model}} & BLEU-4 &      & ROUGE-L &       & METEOR &       \\
\multicolumn{1}{c}{}                           & \multicolumn{1}{c}{}                       & Top1   & Top5 & Top1    & Top5  & Top1   & Top5  \\ \hline

\multirow{6}{*}{Root   cause}                    & Curie                                      & 1.26   & 2.01 & 4.75    & 7.80  & 7.94   & {13.30} \\
                                               & Codex                                      & 1.18   & 1.94 & 3.80    & 7.07  & 6.58   & 12.20 \\
                                               & Davinci                                    & {2.83}   & {4.37} & {6.11}    & {11.55} & 6.04   & 11.87 \\ 
                                               & \bc{Davinci-002}            & 1.35 & 2.5 & 4.89 & 8.58 & 7.65 & {13.55} \\
                                               & \bc{Finetuned-Davinci-002} & \textbf{4.24} & \textbf{7.15} & \textbf{11.43} & \textbf{17.2} & \textbf{10.42} & \textbf{16.8}\\
                                               & \bc{\% gain for Finetuning} & 49.82 & 63.62 & 87.07 & 48.92 & 31.23 & 23.99 \\
                                               \hline

\multirow{6}{*}{Mitigation}                  & Curie                                      & 0.81   & 1.50 & 2.45    & 4.59  & 5.33   & 9.40  \\
                                               & Codex                                      & 0.80   & 1.57 & 1.97    & 4.05  & 4.56   & 8.55  \\
                                               & Davinci                                    & 2.18   & 3.67 & 3.84    & 7.84  & 4.99   & 10.44 \\ 
                                               & \bc{Davinci-002}         & 0.92 & 1.89 & 2.31 & 4.52 & 4.92 & 9.2 \\
                                               & \bc{Finetuned-Davinci-002} & \textbf{6.76} & \textbf{11.66} & \textbf{10.22} & \textbf{18.14} & \textbf{8.23} & \textbf{15.13} \\
                                               & \bc{\% gain for Finetuning} & 210.1 & 217.7 & 166.2 & 131.4 & 54.4 & 44.9 \\ \hline

\end{tabular}
}
\vspace{-0.1in}
\end{table}

\subsection{Does multi-task learning improve the performance of GPT-3.x models at finding root causes and mitigation plans? (RQ4)}
To evaluate the results of multi-task training in the root cause recommendation and mitigating planning tasks, we combine the training set of the two tasks for GPT-3.x models. The models are then individually tested using the corresponding test sets. Table \ref{tbl:mt_rc} shows the results of root cause and mitigation with multi-task training. Overall, we observe that multi-task training does not significantly outperform training for a single task. 
The performance of Curie and Codex models has fallen by an average of 2.8\% for BLEU-4, 2.0\% for Rouge-L and 7.2\% for Meteor. Only the Davinci model is marginally 6.2\% better than single task training in terms of BLEU-4 metric. The performance of Code-davinci-002 is almost always lower across all lexical metrics in a multi-task setting. Similar to this, the results of mitigation generation reveals a 4.1\% performance decline in average for all the four models. The lack of connection between the root cause and mitigation is what mostly contributes to the decline in performance. It is challenging to transfer knowledge from one task to the other because of the distinct distribution in their answer spaces, such as the variations in root cause and mitigation length and concreteness.


\begin{table}[t]

\centering
\caption{Effectiveness of multi-task learning}
\vspace{-0.05in}
\label{tbl:mt_rc}

\resizebox{.95\columnwidth}{!}{%
\renewcommand{\arraystretch}{1.2}

\begin{tabular}{p{1.0cm}p{1.1cm}p{1.0cm}cccccc}

\hline
\multicolumn{1}{c}{\multirow{2}{*}{Objective}}                        & \multicolumn{1}{c}{\multirow{2}{*}{Model}} & \multicolumn{1}{c}{\multirow{2}{*}{\begin{tabular}[c]{@{}c@{}}Multi-\\ tasking?\end{tabular}}} & \multicolumn{2}{c}{BLEU-4} & \multicolumn{2}{c}{ROUGE-L} & \multicolumn{2}{c}{METEOR} \\ 
\multicolumn{1}{c}{}                                                  & \multicolumn{1}{c}{}                       & \multicolumn{1}{c}{}                                                                           & Top1        & Top5         & Top1         & Top5         & Top1        & Top5         \\ \hline
\multirow{8}{*}{\begin{tabular}[c]{@{}l@{}}Root\\ Cause\end{tabular}} & \multirow{2}{*}{Curie}                     & No                                                                                             & 3.40        & 6.29         & 9.04         & 15.44        & 7.21        & 13.65        \\
                                                                      &                                            & Yes                                                                                            & 3.30        & 6.13         & 8.66         & 15.51        & 6.60        & 12.97        \\
                                                                      & \multirow{2}{*}{Codex}                     & No                                                                                             & 3.44        & 6.25         & 8.98         & 15.51        & {7.33}        & {13.82}        \\
                                                                      &                                            & Yes                                                                                            & 3.42        & 6.11         & 8.64         & 15.24        & 6.53        & 12.81        \\
                                                                      & \multirow{2}{*}{Davinci}                   & No                                                                                             & 3.34        & 5.94         & 8.53         & 15.10        & 6.67        & 12.95        \\
                                                                      &                                            & Yes                                                                                            & {3.60}        & 6.27         & {9.11}         & {15.66}        & 7.31        & 13.64        \\
                                                                  & \multirow{2}{*}{\bc{Davinci-002}}                   & \bc{No}  & \textbf{4.24} & \textbf{7.15} & \textbf{11.43} & \textbf{17.2} & \textbf{10.42} & \textbf{16.8} \\
                                                                      &     & \bc{Yes}  & \textbf{4.24} & 7.09 & 11.32 & 17.14 & 10.32 & 16.34 \\\hline
\multirow{8}{*}{Mitigation}                                           & \multirow{2}{*}{Curie}                     & No                                                                                             & 5.47        & 10.62        & 8.03         & {16.31}        & {6.22}        & 12.75        \\
                                                                      &                                            & Yes                                                                                            & 5.49        & {10.89}        & 7.98         & 16.14        & 5.92        & 12.54        \\
                                                                      & \multirow{2}{*}{Codex}                     & No                                                                                             & 5.53        & 10.62        & {8.15}         & 16.23        & 6.19        & {13.15}        \\
                                                                      &                                            & Yes                                                                                            & 5.15        & 10.88        & 7.49         & 15.87        & 5.55        & 11.85        \\
                                                                      & \multirow{2}{*}{Davinci}                   & No                                                                                             & 5.54        & 10.66        & 8.10         & 15.96        & 6.18        & 12.49        \\
                                                                      &                                            & Yes                                                                                            & {5.64}        & 10.74        & 7.88         & 15.97        & 6.13        & 12.99 \\                                                                   & \multirow{2}{*}{\bc{Davinci-002}}                   & \bc{No}  & \textbf{6.76} & \textbf{11.66} & \textbf{10.22} & \textbf{18.14} & \textbf{8.23} & \textbf{15.13} \\
                                                                      &     & \bc{Yes}  & 6.58 & 11.36 & 10.04 & 17.76 & 7.91 & 14.36 \\ \hline       
\end{tabular}
}
\vspace{-0.1in}
\end{table}

\subsection{Do GPT-3.x models get better at proposing mitigation plans if the root cause is given? (RQ5)}
We assess the performance of the mitigation generation while the root cause is being revealed. Our training set of mitigation is reduced from 5,455 to 2,973 as a result of the missing root causes in the incidents, and we have 166 test samples to evaluate the model. Despite the sample reduction in the training set, Table \ref{tbl:mit_rc} reveals a considerable performance gain with the additional root cause information: the average for all three metrics is improved by 9.8\% for the Curie model, 8.3\% for the Codex model, 5.4\% for the Davinci model and 26\% for the Code-davinci-002. Nevertheless, we observe that the performance gain of the Code-davinci-002 model's Top-5 recommendations is modest compared to the improvement of the Top-1 results. Despite this, the overall promising results highlight the significance of root cause information in generating mitigation plans.


\begin{table}[t]

\centering
\vspace{0.1in}
\caption{Effectiveness of GPT-3 models at proposing mitigation plans given root causes}
\vspace{-0.05in}
\label{tbl:mit_rc}

\resizebox{.95\columnwidth}{!}{%
\renewcommand{\arraystretch}{1.2}

\begin{tabular}{llcccccc}
\hline
\multicolumn{1}{c}{\multirow{2}{*}{Model}} & \multicolumn{1}{c}{\multirow{2}{*}{\begin{tabular}[c]{@{}c@{}}Root-cause   \\ given?\end{tabular}}} & \multicolumn{2}{c}{BLEU-4} & \multicolumn{2}{c}{ROUGE-L} & \multicolumn{2}{c}{METEOR} \\
\multicolumn{1}{c}{}                       & \multicolumn{1}{c}{}                                                                                & Top1        & Top5         & Top1         & Top5         & Top1        & Top5         \\ \hline
\multirow{2}{*}{Curie}                     & No                                                                                                  & 5.92        & 11.29        & 9.46         & 17.76        & 7.34        & 13.35        \\
                                           & Yes                                                                                                 & 6.59        & {12.40}        & {10.25}        & 18.61        & 8.24        & 16.00        \\
\multirow{2}{*}{Codex}                     & No                                                                                                  & 6.25        & 11.23        & 8.94         & 17.62        & 6.46        & 13.00        \\
                                           & Yes                                                                                                 & 6.23        & 12.03        & 9.32         & {18.48}        & 7.73        & 15.96        \\
\multirow{2}{*}{Davinci}                   & No                                                                                                  & 6.35        & 12.05        & 8.75         & 18.21        & 7.28        & 15.07        \\
                                           & Yes                                                                                                 & {7.02}        & 11.47        & 9.49         & 18.20        & {8.40}        & {16.17} \\ 
\multirow{3}{*}{\bc{Davinci-002}}                   & \bc{No}  & 6.8 & 12 & 9.48 & 17.37 & 8.15 & 15.53 \\
& \bc{Yes} & \textbf{8.6} & \textbf{13.28} & \textbf{11.56} & \textbf{19.46} & \textbf{10.9} & \textbf{18.08} \\ 
& \bc{\%gain} & 26.47 & 10.21 & 21.94 & 12.03 & 33.74 & 16.42\\   \hline       
\end{tabular}
}
\vspace{-0.1in}
\end{table}

\subsection{Do the models better propose mitigation plans for
machine-detected incidents than human-detected ones? (RQ6)}

\begin{table}[t]

\centering
\caption{Models' performance on machine vs human detected incidents}
\vspace{-0.05in}
\label{tbl:machine_human}

\resizebox{.95\columnwidth}{!}{%
\renewcommand{\arraystretch}{1.2}

\begin{tabular}{llcccccc}
\hline
\multicolumn{1}{c}{\multirow{3}{*}{Model}} & \multicolumn{1}{c}{\multirow{3}{*}{\begin{tabular}[c]{@{}c@{}}Machine\\detected?\end{tabular}}} & \multicolumn{2}{c}{\multirow{2}{*}{BLEU-4}} & \multicolumn{2}{c}{\multirow{2}{*}{ROUGE-L}} & \multicolumn{2}{c}{\multirow{2}{*}{METEOR}} \\
\multicolumn{1}{c}{}                       & \multicolumn{1}{c}{}                                                                                   & \multicolumn{2}{c}{}                        & \multicolumn{2}{c}{}                          & \multicolumn{2}{c}{}                        \\
\multicolumn{1}{c}{}                       & \multicolumn{1}{c}{}                                                                                   & Top1                 & Top5                 & Top1                  & Top5                  & Top1                 & Top5                 \\ \hline
\multirow{2}{*}{Curie}                     & Yes                                                                                                    & 5.49                 & 10.54                & 8.54                  & 16.63                 & 6.45                 & 13.13                \\
                                           & No                                                                                                     & 5.45                 & 10.65                & 7.78                  & 16.15                 & 6.10                 & 12.56                \\
\multirow{2}{*}{Codex}                     & Yes                                                                                                    & {5.76}                 & 10.54                & {9.10}                  & {16.84}                 & {6.80}                 & {13.88}                \\
                                           & No                                                                                                     & 5.41                 & 10.67                & 7.68                  & 15.93                 & 5.88                 & 12.78                \\
\multirow{2}{*}{Davinci}                   & Yes                                                                                                    & 5.56                 & 10.51                & 8.49                  & 16.17                 & 6.34                 & 12.59                \\
                                           & No                                                                                                     & 5.52                 & {10.74}                & 7.91                  & 15.86                 & 5.95                 & 12.44   \\ 
\multirow{3}{*}{\bc{Davinci-002}}                   & \bc{Yes}  & \textbf{7.18} & \textbf{11.83} & \textbf{11.5} & \textbf{18.59} & \textbf{9.41} & \textbf{15.66} \\
& \bc{No} & 6.56 & 11.57 & 9.58 & 17.92 & 7.65 & 14.87 \\ 
& \bc{\%gain} & 9.45 & 2.25 & 20.04 & 3.74 & 23.01 & 5.31 \\   \hline             
\end{tabular}
}
\vspace{-0.1in}
\end{table}

We analyze the mitigation generation performance of GPT-3.x models for both machine and human detected incidents in Table \ref{tbl:machine_human}. We employ the same training set but separate the test samples by the categories of human and machine detected incidents. The testing samples consist of 592 incidents recognized by machines and 1188 incidents detected by humans. Table \ref{tbl:machine_human} demonstrates that machine-recognized incidents can outperform those detected by humans by a factor of 9.5\% for BLEU-4, 20\% for ROUGE-L and 23\% for METEOR in the context of Top-1 recommendations of Code-davinci-002 model. It is due to the fact that machine detected incidents usually adhere to certain patterns, which are easier for machine learning models to recognize.

\section{Looking through the Incident Owners' eyes}
\label{hstudy}

\begin{table*}[h!t]
\centering
\caption{Correctness and readability scores assigned by the incident owners}
\vspace{-0.05in}
\label{tbl:hstudy}
\resizebox{.95\textwidth}{!}{%
\renewcommand{\arraystretch}{1.05}
\begin{tabular}{llcccccccccccccc}
\hline
\multicolumn{1}{c}{\multirow{2}{*}{Objective}} & \multicolumn{1}{c}{\multirow{2}{*}{Criteria}} & \multicolumn{2}{c}{RoBERTA} & \multicolumn{2}{c}{CodeBERT} & \multicolumn{2}{c}{Curie} & \multicolumn{2}{c}{Codex} & \multicolumn{2}{c}{Davinci} & \multicolumn{2}{c}{Davinci-002} & \multicolumn{2}{c}{\begin{tabular}[c]{@{}c@{}}Max\\  OpenAI\end{tabular}} \\
\multicolumn{1}{c}{} & \multicolumn{1}{c}{} & Mean & Median & Mean  & Median & Mean & Median & Mean & Median & Mean & Median & Mean & Median & Mean & Median \\ \hline
\multirow{2}{*}{Root cause} & Correctness & 1.56 & 1 & 1.72 & 1 & 2.40 & 2 & 2.40 & 2 & 2.88 & \textbf{3} & 2.56 & 2 & \textbf{3.52} & \textbf{3} \\
 & Readability & 3.56 & \textbf{5} & 3.68 & \textbf{5} & 3.08 & 4 & 3.52 & 4 & 3.56 & 5 & 3.8 & 4 & \textbf{4.52} & \textbf{5} \\
\multirow{2}{*}{Mitigation} & Correctness & 1.6 & 1 & 1.52 & 1 & 2.28 & 2 & 2.28 & 1 & 3.04 & 3 & 3.16 & 3 & \textbf{4.04} & \textbf{4}\\
 & Readability & 2.88 & 2 & 3.04 & 4 & 2.52 & 2 & 2.8 & 3 & 3.52 & 4  & 4.08 & 4 & \textbf{4.64} & \textbf{5} \\ \hline                               
\end{tabular}
}
\vspace{-0.1in}
\end{table*}

\subsection{Methodology}
From our test sets for root causes and mitigation plans, we selected the incidents with both root causes and mitigation, so that each incident owner could evaluate both the models in the same interview. Incident resolution is a complex task requiring significant context and domain knowledge about the service and also about the specific incidents. Hence, we conducted this human evaluation with the actual owners who root caused and mitigated the incidents. We chose 50 recent incidents which occurred in the last two months, to evaluate the models' performance so that the incident owners could precisely remember what happened during managing particular incidents. We reached out to all the incident owners and 25 incident owners responded and each interview took around 20-30 minutes. 

We presented the outputs from all the models under consideration. For both root causes and mitigation plans, we have six pools of candidates. The first four pools are for OpenAI models, each with six options (including ``none''), and the last two are for RoBERTa and CodeBERT, which has only one candidate. For the OpenAI models, we ask the OCEs to select the best option that might be relevant to the incident. After that, we ask the OCEs to assign correctness and readability for the chosen candidate on a scale of 1-5, with 5 being the best score. Please note that for RoBERTa and CodeBERT, we only have one option. Hence, we only ask to assign correctness and readability scores to those candidates. We define correctness and readability as follows:         

\noindent{\underline{\em Correctness:}} For this metric, we ask the incident owner to check whether the model provides a helpful and relevant suggestion compared to the actual root cause/mitigation. 

\noindent{\underline{\em Readability:}} Readability is the ease with which a reader can understand a generated text. A text is readable if it is grammatically correct, meaningful and easy to understand. Note that a readable text does not need to be correct. 

At the end, we asked the incident owners to assign an overall score (1-5) indicating their perception about the usefulness of LLMs for incident resolution and, also, asked them to share their thoughts and comments regarding this.       
     
\subsection{Results} 
Table~\ref{tbl:hstudy} presents the correctness and readability scores assigned by the incident owners. 
We can see that candidates from the Davinci and Code-davinci-002 pools have achieved higher mean correctness scores than those selected from Curie and Codex models for both root causes (2.88 and 2.56) and mitigation plans (3.04 and 3.16). The mean readability score ranges from 2.52 to 4.08 for all the models. The incident owners expressed positive opinions about the readability of the outputs, and all the models achieved higher readability than correctness scores. We received a few recommendations on how to improve the readability in the future (\eg avoiding use of acronyms and generating more specific or informative comments).

As discussed before, the baseline encoder-decoder models generate very generic comments, and the automatic metrics fail to detect that. We can see the incident owners assign a lower correctness score to RoBERTa and CodeBERT model, 
and several OCEs pointed out the generic nature of the recommendations generated by the encoder-decoder models. Though the correctness score of the OpenAI models ranges from 2.28 to 3.16, several OCEs pointed out that the models recommend beneficial root causes and mitigation plans. For example, the models succeeded in pinpointing  some hard to detect root causes:

\textbf{\textit{``I am very impressed because one model found the right root cause, which was very hard to detect. We found it in the postmortem phase. However, I am a little worried that there would not be enough information on the incident website. Overall, I am impressed with the efficacy of the models.''}}

\textbf{\textit{``Even if not always correct, these suggestions can guide the OCE towards actual root cause. ML model can give directions and can be valuable suggestions.''}}

We also took the maximum score assigned by the OpenAI models and reported the average correctness and readability score. The mean correctness and readability score ranges from 3.52 to 4.64 (median score 3-5), presenting the overall strength of the models. We asked for the overall scores (1-5), and Table~\ref{pie} shows that the incident owners found the overall contribution promising and useful. More than 70\% of incident owners gave three or above for the recommendations of the models. We found that at least one model is effective for most incidents. We also found out why the automatic metrics fail to provide valuable insights. 

There is always another side to the coin, and we observe that the models' outputs are not helpful for some incidents. The OCEs assigned lower scores to those incidents and here are some of the concerns they mentioned:

\textbf{\textit{``Based on just incident data it is difficult for the model to predict root-cause and mitigation because not all data are recorded in the database and some of them are classified.''}}

\textbf{\textit{``Major concern is if the suggestion is incorrect, on-call engineers may take longer time to investigate the problem.''}}

We observed some negative samples for the model because a lack of discussion or other information results in the deprivation of valuable signals from the input. However, the model's overall performance is quite promising, which can be considered a stepping stone toward the automation of root causes and mitigation plans in the future.

\begin{table}[ht]

\centering
\vspace{-0.05in}
\caption{Usefulness of LLMs for incident resolution}
\vspace{-0.05in}
\label{pie}

\resizebox{.5\columnwidth}{!}{%
\renewcommand{\arraystretch}{1.2}

\begin{tabular}{ccc}
\hline
Score & \begin{tabular}[c]{@{}c@{}}\# of incident   \\ owners\end{tabular} & \begin{tabular}[c]{@{}c@{}}In percent (\%)   \\ of total\end{tabular} \\ \hline
5     & 2                                                                  & 7.41                                                                  \\
4     & \textbf{9}                                                                 & \textbf{33.33}                                                                 \\
3     & 8                                                                  & 29.63                                                                 \\
2     & 6                                                                  & 22.22                                                                 \\
1     & 2                                                                  & 7.41 \\ \hline                                                                 
\end{tabular}
}
\vspace{-0.05in}
\end{table}

\subsection{Two illustrative examples}
Table~\ref{example1} exhibits two samples to show the effectiveness of GPT-3.x model for generating root causes and mitigation plans in cloud incidents. We present the actual texts written by the OCEs and generated texts by one of the models (fine-tuned Code-davinci-002) side by side. Though the human-written and generated texts are different, the generated texts provide very relevant and valuable suggestions that can help the OCEs. These textual dissimilarities also show why the OCEs found the GPT-3.x models promising, but the automatic evaluation metrics failed to do that.

\begin{table*}[ht]

\centering
\vspace{-0.05in}
\caption{Examples exhibiting the effectiveness of GPT-3.x model for incident resolution}
\vspace{-0.05in}
\label{example1}

\resizebox{.95\textwidth}{!}{%
\renewcommand{\arraystretch}{1.2}
\begin{tabular}{llll}
\hline
\multicolumn{1}{c}{Examples} & \multicolumn{1}{c}{Task} & \multicolumn{1}{c}{Original text}                                                                                                                                                                                                 & \multicolumn{1}{c}{Generated by DaVinci model}                                                                                            \\ \hline
\multirow{2}{*}{I}             & Root cause                & \begin{tabular}[c]{@{}l@{}}The customer was sending very large amounts of data as their payload. \\ The size in addition to the rate at which these transactions were being \\ sent led to the initial network crash.\end{tabular} & \begin{tabular}[c]{@{}l@{}}Large payloads that exceed the max payload size limit of the ledger \\ caused the ledger to crash.\end{tabular} \\
                               & Mitigation                & A code upgrade with higher payload size solved the problem.                                                                                                                                                                        & Republished the ledger with a higher payload size                                                                                          \\ \hline
\multirow{2}{*}{II}            & Root cause                & \begin{tabular}[c]{@{}l@{}}An unexpected power loss caused multiple VMs to go down in East \\ US causing DB services.\end{tabular}                                                                                            & Datacenter outage impacting multiple services                                                                                              \\
                               & Mitigation                & DB team mitigated the upstream issue in the West US region                                                                                                                                                                         & DB team mitigated the issue by restarting the service.                                                                                 \\ \hline 
\end{tabular}

}
\vspace{-0.05in}
\end{table*}
\section{Discussion \& Threats}


\subsection{Do automatic metrics reflect human perception?}

Automatic evaluation metrics are known to be representative of human perception and are widely used in problems like natural language translation~\cite{vaswani2017attention,lewis2019bart,raffel2019exploring}. Though some recent works looked into the effectiveness of these metrics in code summarization and reported many pitfalls and weaknesses of these metrics~\cite{shia2022evaluation,roy2021reassessing,gros2020code,haque2022semantic}, researchers are still using them for benchmarking. The best possible alternative to automatic metrics is human validation or some form of automatic test case evaluation (done in code generation tasks). The main challenge in incident management is that even experts face difficulties evaluating the incidents if they are not involved in resolving particular incidents. In some cases, the OCEs could not clearly remember the incidents if they happened two months ago. Thus conducting a large-scale study is quite challenging in this area. However, we interviewed 25 incident owners and found that the models perform pretty well even after achieving lower scores with automatic metrics. We calculated the Pearson coefficient for all three lexical metrics (\ie BLEU-4, ROUGE-L, and METEOR) with the correctness and readability score assigned by the OCEs. We observed that the co-efficient varies from -0.42 to +0.62, preventing us from getting specific patterns in the value. 
That also indicates that these automatic metrics may not be coherent with human perception for resolving cloud incidents. However, more sample cases are needed to reach any concrete resolution.  

\subsection{Natural language or code? Which family of models are better for incident management?}
While choosing the models, we selected both natural language (\ie RoBERTa, Curie, Davinci) and code models (\ie CodeBERT, Codex-cushman, Code-davinci-002) to see which family of models is beneficial for incident management. We did not find any winners from these two groups. Davinci and Code-davinci-002 models are found to be producing correct and readable suggestions compared to other models. Note that both of them have 175 billion parameters. We leave fine-tuning larger code models or pre-training a model from scratch with incident data for future research.

\subsection{How the models' performance can be improved?}
We received several recommendations from the incident owners. The main recommendation is to incorporate the discussions among the OCEs into the model. This will guide the model to locate better suggestions. We also dropped many incidents with summaries that written or updated at the time of incident resolution. To fairly evaluate the model and prevent possible data leakage (root cause and mitigation can be written in summary if updated later), we discarded them from our dataset. Incorporating them into our dataset after preventing data leakage may improve the performance of the models. We also lost some critical information while cleaning the summaries (\eg discarding images and tables). Incorporating that information may also help.         

\subsection{Threats to Validity}
There are several threats to our study. The semantic metrics use pre-trained models at the core, and we use the default, natural language models for the evaluation. A model pre-trained with incident management text may result in some changes in the performance evaluation. Also, we train and evaluate the models with the services available within our organization. These models may show unexpected behaviors if evaluated on a different set of services from other organizations. Some incidents owners expressed concerns about the models' efficacy with rare incidents, and rare incidents are frequently reported at \company. Another threat to our study is the sample size of our human subject study. It is difficult to achieve statistical significance on correctness and readability scores with such small samples. However, it is challenging to scale depending on the nature of the study. 
\section{Related Work}

\subsection{Incident management}
Incident management in large cloud services has become a popular topic of research in the Systems and Software Engineering communities. Prior work in this space has focused on two main directions. First, there has been several empirical studies on analyzing incidents and outages in production systems which have focused on studying incidents caused by certain type of issues \cite{leesatapornwongsa2016taxdc, alquraan2018analysis, gao2018empirical, zhang2021understanding} or issues from specific services and systems \cite{ghosh2022fight, liu2019bugs, yuan2014simple}. Second and more related to our work is the use of machine learning and data driven techniques for automating different aspects of incident life-cycle such as triaging \cite{EmpiricalIcMICSE2019, ContinuousTriageASE2019, azad2022picking}, diagnosis \cite{nair2015learning, bansal2019decaf, luo2014correlating} and mitigation \cite{jiang2020mitigate}. Different from prior work, this is the first effort on leveraging state-of-the art language models for assisting OCEs with incident resolution. We hope that this work will also motivate future work which will merge traditional task-specific discriminative models with LLMs to do end-to-end automation of production incidents.

\subsection{LLMs in Software Engineering}
Even though this is the first work leveraging LLMs for AIOps, several works in Software Engineering have tried to solve other challenging problems with LLMs. Github Copilot uses GPT-3 for automated code generation from natural language inputs~\cite{chen2021evaluating}. Several researchers have addressed code generation~\cite{chen2021evaluating,xu2022systematic}, docstring generation~\cite{chen2021evaluating,ahmed2022few}, and code repair~\cite{fan2022improving,joshi2022repair} problems. Barei{\ss} \etal~\cite{bareiss2022code} show how few-shot learning can be effective at (i) code mutation; (ii) test oracle generation from natural language documentation; and (iii) test case generation task. Jain \etal propose an approach to augment large language models with post-processing steps based on program analysis and synthesis techniques and achieve better performance~\cite{jain2022jigsaw}. However, unlike code generation where we have both lexical and structural information along with massive amount of training data, we explore the problem of incident resolution using state-of-the-art LLMs which has not been done before.

\section{Conclusion}

With this work, we show that state-of-the-art large language models such as GPT-3 and GPT-3.5 are effective to help with incident management, specifically, to identify root causes and mitigation steps. To compare the effectiveness of the models, we conducted a rigorous and large-scale study at \company{}, on over 40,000 incidents. To assess the actual usefulness of the approach, we involved the actual owners of production incidents. We expect that this paper is the first of many studies that leverage LLMs to make incident management more effective. 
Our next steps are to deploy the models in production to assist the OCEs with incident resolution. We are also planning to explore other usage scenarios for LLMs such as incident summarization.

\section{Acknowledgements}
We would like to thank the engineers who participated in the validation of root causes and mitigation steps. We would like to also acknowledge the contributions of the following people across Microsoft: Oleg Losinets, Jim Jernigan and Jim Kleewein.

\bibliographystyle{IEEEtran}
\bibliography{incident.bib}

\end{document}